\documentclass[11pt,a4paper,onecolumn,usenatbib]{article}
\usepackage{psfig,natbib}
\usepackage{graphics,graphicx}

\usepackage{helvet}

\begin{document}

\begin{center}
\section*{\bf {\sc {\LARGE The Kinematics of Star Formation:\\Theory and Observation in the Gaia Era}}}

{\large A summary of the RAS Specialist Discussion Meeting held in January 2015}\\
\vspace{0.5cm}
Nicholas J. Wright, Keele University\\
\end{center}
\vspace{0.5cm}

The European Space Agency's Gaia space telescope, launched in 2013, aims to measure the positions, parallaxes, and proper motions of a billion stars in our Galaxy and throughout the Local Group. In doing so it will include hundreds of thousands of young stars in star forming regions, star clusters and OB associations. This will provide us with an unprecedented view of the role of dynamics in the star formation process and the evolution of young star clusters. Data from this ambitious mission is expected very soon, with the first data release scheduled for 2016. This review discusses the current state of our understanding of the kinematics of star formation and how the community can best prepare for Gaia data.

\section*{Introduction}

Stellar dynamics plays a crucial role, both in the star formation process and in the formation and evolution of star clusters. The motions of stars controls the evolution of star-forming structures, the formation and dissolution of star clusters, and the level of dynamical interactions between stars in dense stellar systems. This is particularly critical during the early years of a star's life whilst proto-planetary disks are evolving and planetary systems forming. Stellar kinematics is also crucial for our understanding of the star formation process, allowing us to distinguish between dynamic and quasi-equilibrium theories of star formation, to trace back the motions of young stars to their birth sites, and to test claims of isolated massive star formation. However, our knowledge of the kinematics of young stars and protostars is very limited. The recent increase in kinematic data from large radial velocity surveys has improved the situation, but we are now on the brink of a step change in stellar dynamics with the imminent release of data from the Gaia satellite.

Gaia \citep{perr01} is an ESA mission that was launched in 2013 with full operations beginning in 2014. It is the successor to the Hipparcos mission, the first precision-astrometry space mission, which produced a catalog of positions, parallaxes and proper motions for more than 100,000 stars. Gaia, which is part of ESA's Horizon 2000+ long-term scientific programme, will take this further by targeting a billion stars across the Milky Way.

Gaia's payload consists of a single integrated instrument with three major functions: astrometry, photometry, and a radial velocity spectrometer. All the instruments share the two telescopes and a common focal plane, albeit with different areas for the different functions of the instrument. The astrometric instrument will precisely determine the positions of nearly all stars down to $G = 20^{th}$ magnitude (Gaia's $G$ band extends from $\sim$400 to 900~nm) approximately 70 times over the 5~year lifetime of the mission. This data will be used to calculate 5-parameter astrometric solutions for all stars, including positions, proper motions and parallaxes. The end-of-mission proper motions will reach a precision of $< 10$~$\mu$as/yr for stars with $G < 14$~mag and $< 100$~$\mu$as/yr for $G < 18.5$~mag. At the distance of the Orion Nebula Cluster for example this is equivalent to velocity uncertainties of $< 0.02$~km/s ($G < 14$~mag) and $< 0.2$~km/s ($G < 18.5$~mag).

Gaia's main goal is to construct the first 3 dimensional map of astronomical objects throughout the Milky Way, thereby revealing the structure, dynamics, evolutionary history and formation of our Galaxy. In addition to this Gaia is expected to detect tens of thousands of Jupiter-sized planets, hundreds of thousands of quasars, and contribute to many other areas of astrophysics. It will also make a significant impact on studies of star formation and star clusters, providing parallaxes and kinematics for hundreds of thousands of young stars in a wide variety of environments. In addition to this a number of current and planned spectroscopic surveys of star forming regions and star clusters (e.g., IN-SYNC and the Gaia-ESO Survey) are providing radial velocities, spectral types, and diagnostics of youth that together will drive a transformational improvement in kinematic data quality. This will soon allow us, for the first time, to resolve cluster kinematics for large numbers of stars in 3D, providing estimates of energies, angular momenta and dynamics and avoiding the need for simple isotropy assumptions.

The RAS Specialist Discussion meeting on 9 January 2015 gathered experts in star formation, star clusters and stellar kinematics from across the UK and Europe to discuss the current state of our understanding of the star formation process and the role of kinematics within it. The motivation for the meeting was to consider what impact Gaia and other kinematic observations will have on this understanding, as well as how the community can best prepare for the data coming from these facilities.

\section*{What role does dynamics play in the star formation process?}

\begin{figure}[t]
\begin{center}
\includegraphics[width=9.2cm]{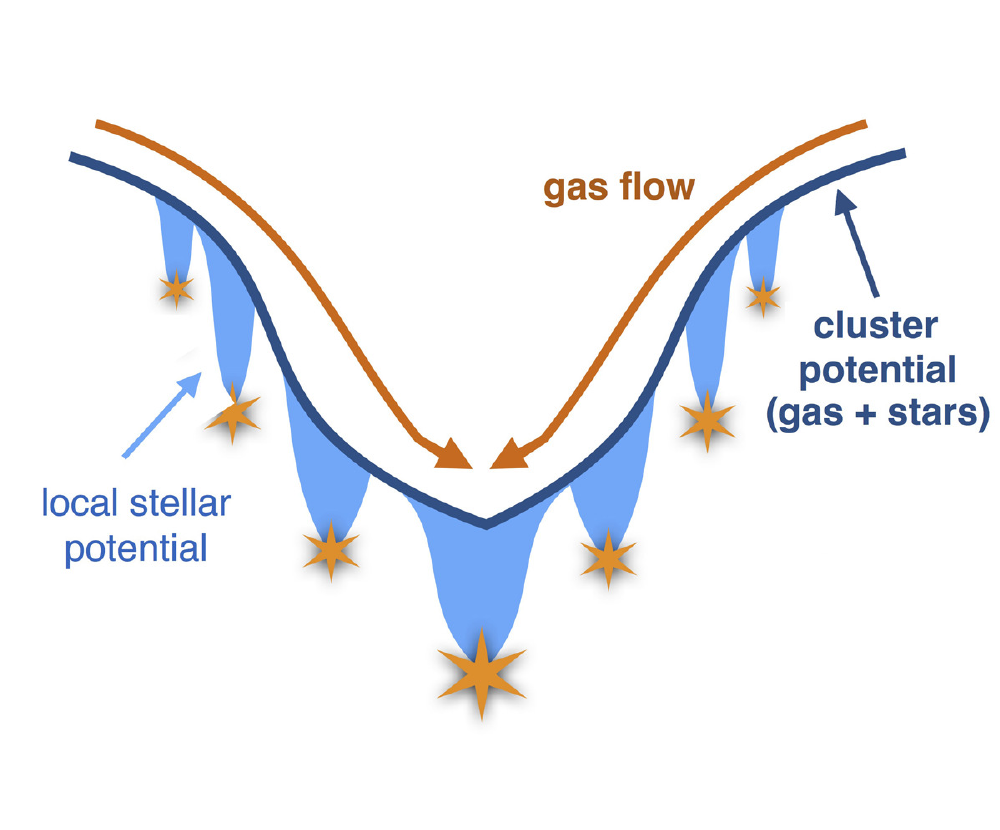}
\includegraphics[width=7.5cm]{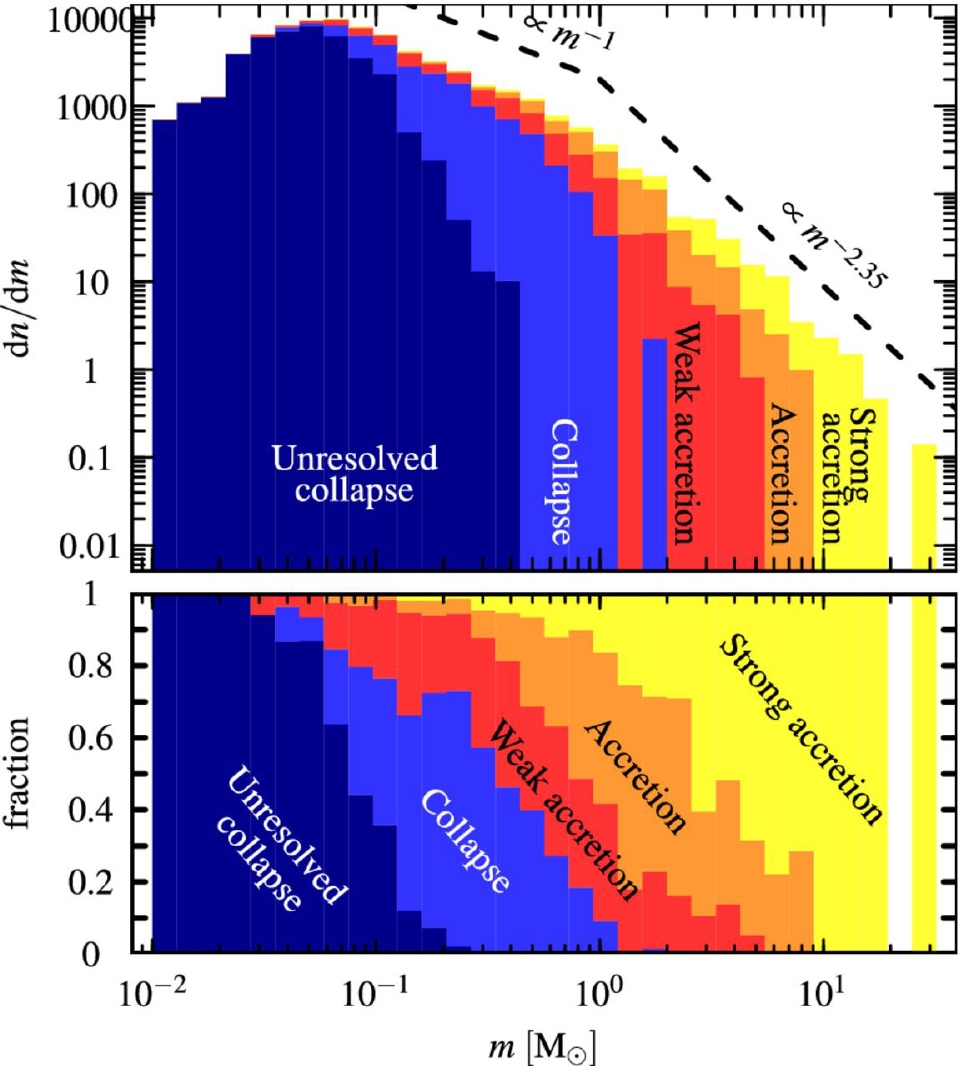}
\caption{\it \small Left: Illustration of the idea behind the theory of competitive accretion \citep{zinn82,bonn01}. The gravitational potential of the forming star cluster, due to both the gas and the stars, funnels gas flow towards the centre of the potential. Gas that flows into the local stellar potentials surrounding the forming stars may then be accreted onto those pre-stellar cores. The cores at the centre of the cluster potential are able to accrete more mass because of their location, so they are more likely to grow to be the most massive stars in the cluster. Image credit: Paul Clark. Right: Histogram of the final sink-particle masses colour-coded by their mass growth histories \citep[see][for details of the classes of mass growth]{masc14}. The amount of accretion that contributes to a sink particle's final mass increases with increasing final sink-particle mass. The top panel shows the total number of sink-particles and the bottom panel shows the fraction of each class within a given mass range. Image (right-hand panel) from \citet[][Figure 9]{masc14}.}
\end{center}
\end{figure}

The first session, chaired by Rowan Smith (University of Manchester), focused on reviewing our current understanding of the star formation process and the role of kinematics within that process. Paul Clark (Cardiff University) gave the first invited review of the day and discussed the current theories that seek to describe the star formation process. Paul started off by discussing the two main pictures of star and cluster formation, competitive accretion and turbulent fragmentation. In the competitive accretion idea \citep{zinn82,bonn01} the gravitational potential of the cluster, due to both stars and gas, funnels gas flow towards the centre of the potential (see Figure~1). The amount of accretion onto each pre-stellar core then depends on the amount of gas surrounding the core that is bound to the core itself. This itself is dependent on the velocity of the gas, the mass of the core, and its position within the overall cluster potential. Pre-stellar cores at the centre of the cluster are able to accrete more mass because of their location at the deepest point in the cluster potential. These cores are therefore more likely to grow to become the most massive stars in the cluster. Competitive accretion therefore predicts that the most massive stars should form at the centres of the gravitational potential wells. Unfortunately observers cannot directly identify the centres of the potential wells and so it is more common to correlate the positions of the massive stars with the areas of the highest stellar density, but as Paul pointed out these are not always the same.

The other key model of star formation is that of turbulent fragmentation, sometimes referred to as core accretion \citep{fede12}. In this model pre-stellar cores are created by supersonic turbulence, but they then decouple from the surrounding gas flows so that the mass of the stellar system that forms is proportional to the mass of the initial pre-stellar core. This idea is supported by the fact that the distribution of core masses, both from theory and observation, has a similar form, albeit shifted to higher masses, to that of the stellar initial mass function \citep[IMF, e.g.,][]{pado02}. The two theories make different predictions for the spatial distribution of stars during the star formation process as well as their dynamics. Competitive accretion is often considered a more dynamic theory of star formation that potentially benefits from a highly clustered environment, while core accretion is thought to form stars in quasi-equilibrium, and is thus capable of acting in relatively low density environments or potentially even in isolation.

Paul then went on to discuss how different simulations seek to explain how the IMF is built up. \citet{masc14} studied the different sources of accretion within a recent hydrodynamical simulation of star formation. They found that while the peak of the IMF is caused by fragmentation and gravitational collapse within the molecular cloud, the Salpeter-like high-mass tail is driven by strong accretion from the surrounding gas (see Figure~1) and is unrelated to classical Bondi-Hoyle accretion. They also found that while half of the stars have their final masses determined from the initial collapse, the other half acquire most of their mass through the extended accretion phase. \citet{clar08a} found that the dynamics of the molecular cloud can also be very important for how stellar masses are built up in simulations of star formation. They found that while gravitationally bound molecular clouds were able to produce IMFs in agreement with the observations, unbound clouds could not sustain competitive accretion and produced much flatter mass functions that those observed.

Finally, Paul outlined how Gaia observations could be used to test our understanding of the star and cluster formation processes. Thanks to its precise proper motions Gaia will be very effective in identifying young stars that have been ejected from various star formation sites, allowing their motions to be traced back to their birth clusters. The properties of these stars will be vital for constraining the dynamics of star forming regions. For example the age distribution of ejected stars may provide constraints on the timescale of star formation, distinguishing between slow \citep{krum05} and rapid \citep{ball99} models of star formation, while the kinematics of the ejected stars may provide constraints on the structure and density of the star forming region.

\begin{figure}[t]
\begin{center}
\includegraphics[width=17cm]{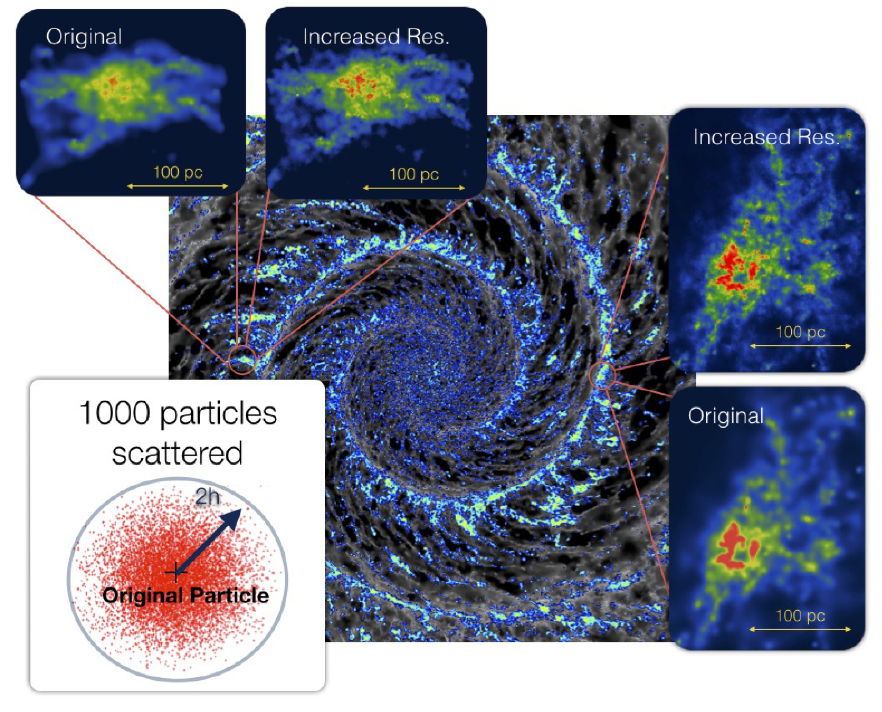}
\label{rey-raposo}
\caption{\it \small Smooth particle hydrodynamical (SPH) simulation of the formation of molecular clouds in a spiral galaxy \citep{dobb13} showing two selected molecular clouds extracted from the simulation with their resolution increased to facilitate more detailed simulations \citep{reyr15}. In the lower left the method for increasing the resolution of the molecular clouds is shown, whereby extra particles are distributed according using the SPH kernel. Image originally from \citet[][Figure 1]{reyr15}.}
\end{center}
\end{figure}

Ram\'on Rey Raposo (University of Exeter) gave the first contributed talk of the day and discussed the need for simulations of the star formation process to take into account the environment of the host galaxy. Stars form in giant molecular clouds (GMCs) that are aggregates of cold and dense molecular clouds distributed across galaxies. The formation of GMCs is thought to be driven by a number of processes including stellar feedback, turbulence, large-scale instabilities, and the agglomeration of smaller clouds. Each of these processes acts over different spatial scales and timescales, and therefore different mechanisms can dominate in different environments within galaxies. To accurately reflect star formation in a galaxy such as our own it is important for simulations to take into account environmental effects such as the external radiation field and the dynamics of the galaxy itself. Most simulations to date use simplified initial conditions such as using turbulent spheres or boxes with a tightly constrained phase space of mass, size and velocity power spectrum.

Ram\'on showed that realistic initial conditions for molecular clouds could be extracted from full scale galactic simulations that include galactic dynamics, chemistry, interstellar medium cooling and heating, and stellar feedback (see Figure~2). Individual molecular clouds in these large-scale simulations can be extracted and used as the initial conditions for more detailed simulations. The resolution of the clouds can be increased by adding extra particles distributed according to the SPH kernel. His results show that the velocity field inherited from the galactic scale simulations can have a major impact on the star formation rate within individual molecular clouds, with clouds that are in the process of collapsing having higher star formation rates that non-collapsing clouds \citep{reyr15}.

\section*{The role of kinematics in the formation of star clusters}

\begin{figure}[t]
\begin{center}
\includegraphics[width=17cm]{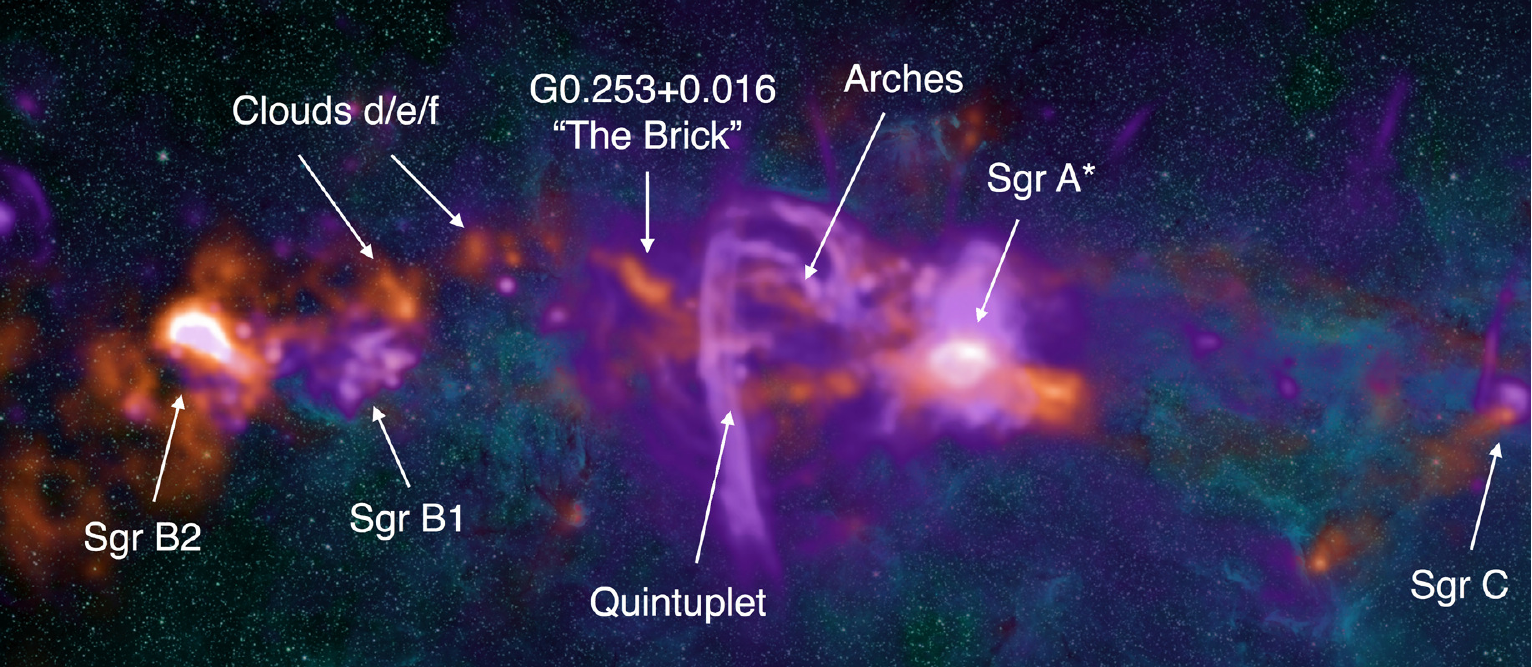}
\label{galactic-centre}
\caption{\it \small Star formation near the Galactic Centre showing the central molecular zone and the common time sequence of star formation that exists from compact GMCs (e.g., G0.253+0.016) to YMCs (e.g., the Arches cluster). Image spans $1.5 \times 0.7$ degrees, or $220 \times 100$~pc at a distance of 8.5~kpc, and is composed of 20cm (purple), 1.1mm (orange), and 8$\mu$m (cyan) observations. Image credit: NRAO / Adam Ginsberg / Jonathan Henshaw.}
\end{center}
\end{figure}

Jonathan Henshaw (Liverpool John Moores University) discussed attempts to understanding the formation of young massive clusters (YMCs) using molecular observations of the large-scale velocity structure in the central molecular zone of our galaxy (see Figure~3). YMCs, such as the Arches and Quintuplet clusters, are rare in our galaxy with only a handful known. It is not currently known whether such massive and dense star clusters form from dense initial conditions, with star formation occurring ``in situ'' (i.e. the stars form in a dense and highly clustered distribution) or whether the stars form at lower densities and then collapse to form a dense cluster \citep{long14}.

The central molecular zone of our galaxy is a prime environment to search for dense molecular cloud precursors, since it is home to two YMCs in the Arches and Quintuplet clusters, and also to the massive ($10^5$~M$_\odot$) and compact ($r \sim 2-3$~pc) GMC G0.253+0.016, commonly known as ``The Brick'', a strong candidate to be a progenitor YMC. All these structures are part of a common time sequence of star formation in which GMCs are squeezed as they pass close to the supermassive black hole Sgr~A*, inducing star formation within them \citep{long13}.

The observations presented by Jonathan show that the velocity structure of the central molecular zone is very complex and dominated by multiple non-thermal components. Traditional moment analysis methods for an analysing such complex data proved inadequate and instead line-fitting was necessary to produce detailed insights into the kinematics. They find evidence for several distinct regions each with their own coherent kinematics. These regions are part of a physically linked gas stream that supports the view of a common timeline of star formation in the vicinity of Sgr~A*. They also find evidence for velocity oscillations in their kinematic data, which may signify either gravitational instabilities or gas flows from surroundings GMCs.

\begin{figure}[t]
\begin{center}
\includegraphics[width=17cm]{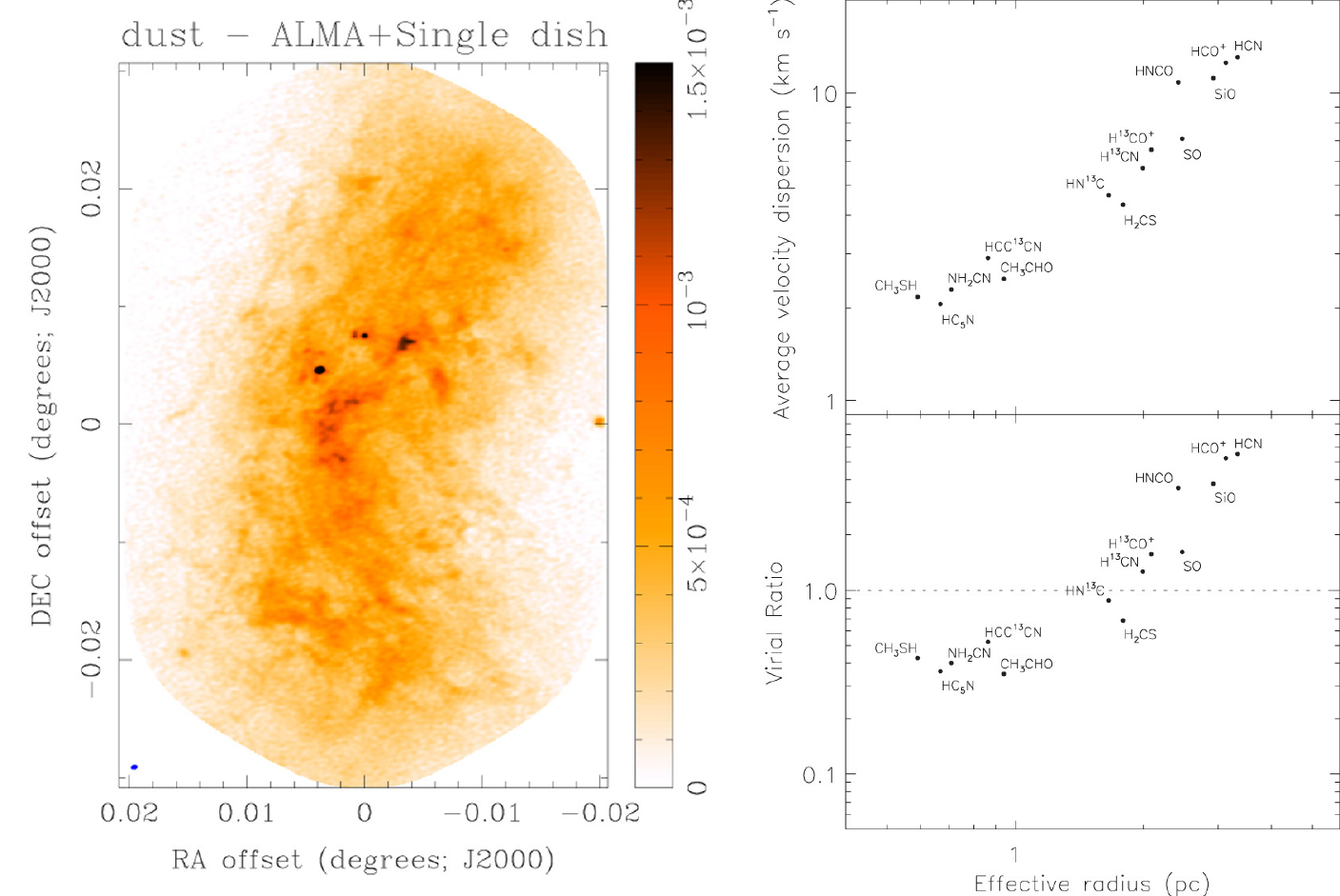}
\label{the-brick}
\caption{\it \small Left: Combined ALMA and single dish 3mm dust emission image of G0.253+0.016, ``The Brick'', showing the detailed structure revealed by these observations. Right: Virial ratio as a function of effective radius from various molecular line transitions observed towards G0.253+0.016 \citep{rath15}. For each molecular line the spatial extent of the emission from that line provides the effective radius, while the velocity dispersion of the emission over that spatial extent provides the velocity dispersion, from which the virial ratio can be calculated with knowledge of the total mass within that radius (estimated using the known radial dust mass). This shows that while the densest parts of the molecular cloud are bound, the outer parts are unbound and expanding. Image from \citet{rath15}, reproduced by permission of the AAS.}
\end{center}
\end{figure}

Nate Bastian (Liverpool John Moores University) talked further about the role of dynamics in the formation of stars and star clusters. Nate started by reminding us of the simple model of early cluster evolution in which dense and compact star clusters are born embedded within molecular clouds, but are gravitationally unbound when feedback disperses the gas left over from star formation \citep[e.g.,][]{lada84}. This theory of cluster disruption, often referred to as {\it residual gas expulsion}, predicts that a large fraction of young and non-embedded star clusters should be supervirial and in the process of expanding. However most kinematic observations of exposed star clusters have shown them to be in (or close to) virial equilibrium \citep[e.g.,][]{roch10,hena12,cott12}. This raises the question of why gas expulsion hasn't influenced the virial state of these clusters?

There are a number of possible explanations that have been considered to explain why residual gas expulsion may not have disrupted these clusters. For example it has been suggested that stars may form with initially sub-virial velocities that might allow the star clusters that form to remain in virial equilibrium even after gas dispersion \citep{offn09,smit11}. Alternatively if the star formation efficiency (the fraction of gas converted into stars) in these clusters was sufficiently high ($>$50\%) then more gas would have been converted into stars during their formation and therefore the cluster may not have become unbound by the removal of the remaining gas potential \citep{good06}. Finally it has also been suggested that stars may naturally become spatially decoupled from the gas from which they formed during the cluster formation process and therefore that the star cluster would not be dependent on the gas potential to remain bound \citep{krui12}. 

To resolve this issue it would be useful to understand the initial conditions for the formation of YMCs, though the rarity of YMC precursors makes this difficult and suggests that their formation must be a very rapid process. A number of candidate YMC progenitors are observed near the Galactic Centre including a GMC known as ``The Brick'' (see Figure~3), which appears to be gravitationally bound on small scales, but unbound on large scales \citep[see Figure~4,][]{rath15}, though star formation has not yet begun. Such GMCs have also been found to have shallower density profiles than the star clusters that form from them, suggesting that either during or directly after star formation these structures experience a phase of rapid contraction \citep{walk15}.

Gaia will provide an opportunity to study these issues in greater detail, as well as address questions such of whether OB associations are expanding star clusters or whether they form with low densities \citep{wrig14b}, and whether massive stars can form in isolated or low-density environments, or if their dynamics can explain their formation elsewhere \citep{bres12}.

\section*{The kinematics of young stars in star clusters and OB associations}

\begin{figure}[t]
\begin{center}
\includegraphics[width=17cm]{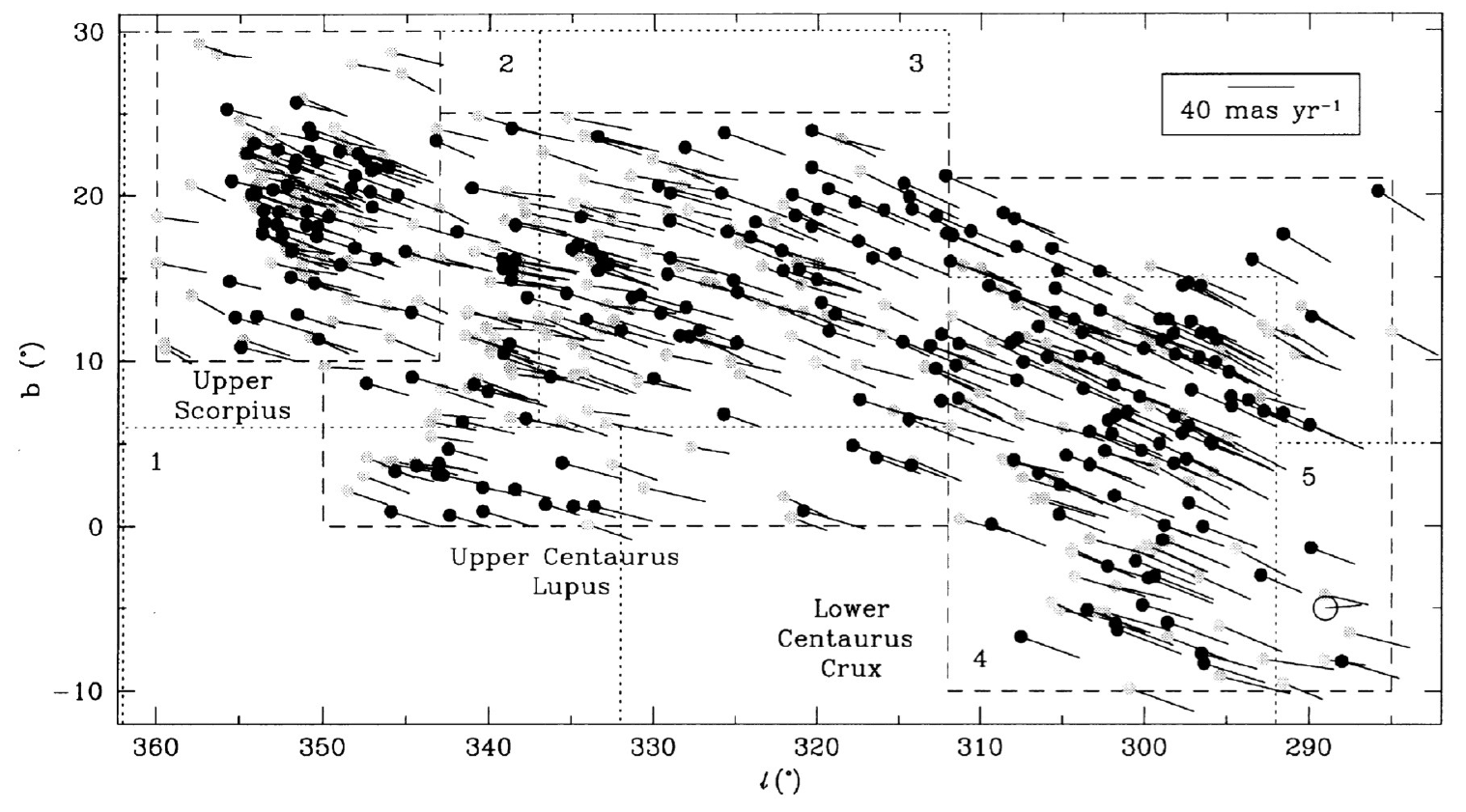}
\label{scorpius-centaurus}
\caption{\it \small The Scorpius-Centaurus association, divided into the subgroups of Upper Scorpius ($\sim$5~Myr), Upper Centaurus-Lupus ($\sim$17~Myr), and Lower Centaurus-Crux ($\sim$16~Myr). At a distance of $\sim$150~pc this OB association has formed almost $10^4$~M$_\odot$ of stars in the recent past, a considerable fraction of all the young stars in the solar vicinity. Image from \citet{deze99}, reproduced by permission of the AAS.}
\end{center}
\end{figure}

The second session, chaired by Richard Parker (Liverpool John Moores University) considered how kinematic observations of young stars in star clusters and OB associations could help constrain the star formation process and improve our understanding of the early evolution of star clusters.

The second invited review of the day was given by Simon Goodwin (University of Sheffield), who raised the question of whether most stars form in dense star clusters or loose OB associations. It is important to constrain the birth environment of stars, particularly the stellar density, because this can have a considerable influence on the early evolution of stars. The amount of time that a star spends at stellar densities $> 100$~pc$^{-3}$ will influence the amount of dynamical interactions that the star will experience with other stars, which in turn affects the evolution of protoplanetary disks, young planetary systems, and stellar multiplicity. Density also plays a critical role in theories of massive star formation, with models such as competitive accretion requiring a dense environment and a large gravitational potential well for stars to accrete sufficient quantities of gas. An important question then is whether star formation is universal and all stars form at either low or high densities, or whether there are multiple modes of star formation in which stars can form in a variety of environments.

A common model of star formation suggests that most stars are born in clusters, but that the majority of these clusters are disrupted due to processes such as residual gas expulsion \citep{lada03}. Observations of star clusters have broadly supported this picture, partly thanks to the ease with which star clusters can be studied. OB associations however are much harder to study, due both to the large area on the sky that they cover and their low stellar density, which can make it difficult to efficiently identify members of the association. Despite this OB associations are an important source of young stars, with the majority of nearby young stars distributed in OB associations \citep{deze99}. For example the Scorpius Centaurus association and its subgroups, at a distance of approximately 150~pc, have formed almost $10^4$~M$_\odot$ of stars, a considerable fraction of all the young stars in the solar vicinity (see Figure~5).

Recent studies have suggested that OB associations may form at low densities \citep{wrig14b}, suggesting both that the massive stars within them must have formed monolithically and not due to competitive accretion, but also that the effects of dynamical encounters are not very important in these environments. If the star formation process is universal, as many believe, then the observation of star formation in a low density environment implies that all stars must form in low density environments. Dense star clusters must therefore form later, potentially from the cold collapse of a low density but sub-virial group of stars \citep{park14}, or possibly under very different and extreme conditions when the star formation rate is much higher. It would seem to be much easier to form a low density OB association than a dense star cluster. The latter requires assembling a large mass of gas into a very dense, cold and gravitationally bound compact clump, within which star formation then occurs, while to form an OB association the gas can be forming stars at much lower densities and over larger areas of space. Answering these questions requires kinematic information for thousands of stars to study the dynamics of star clusters and expanding OB associations, as well as precise distances to help identify the members of these associations. The combined parallaxes and proper motions that Gaia will provide will likely be key to resolving this issue.

\begin{figure}[t]
\begin{center}
\includegraphics[width=17cm]{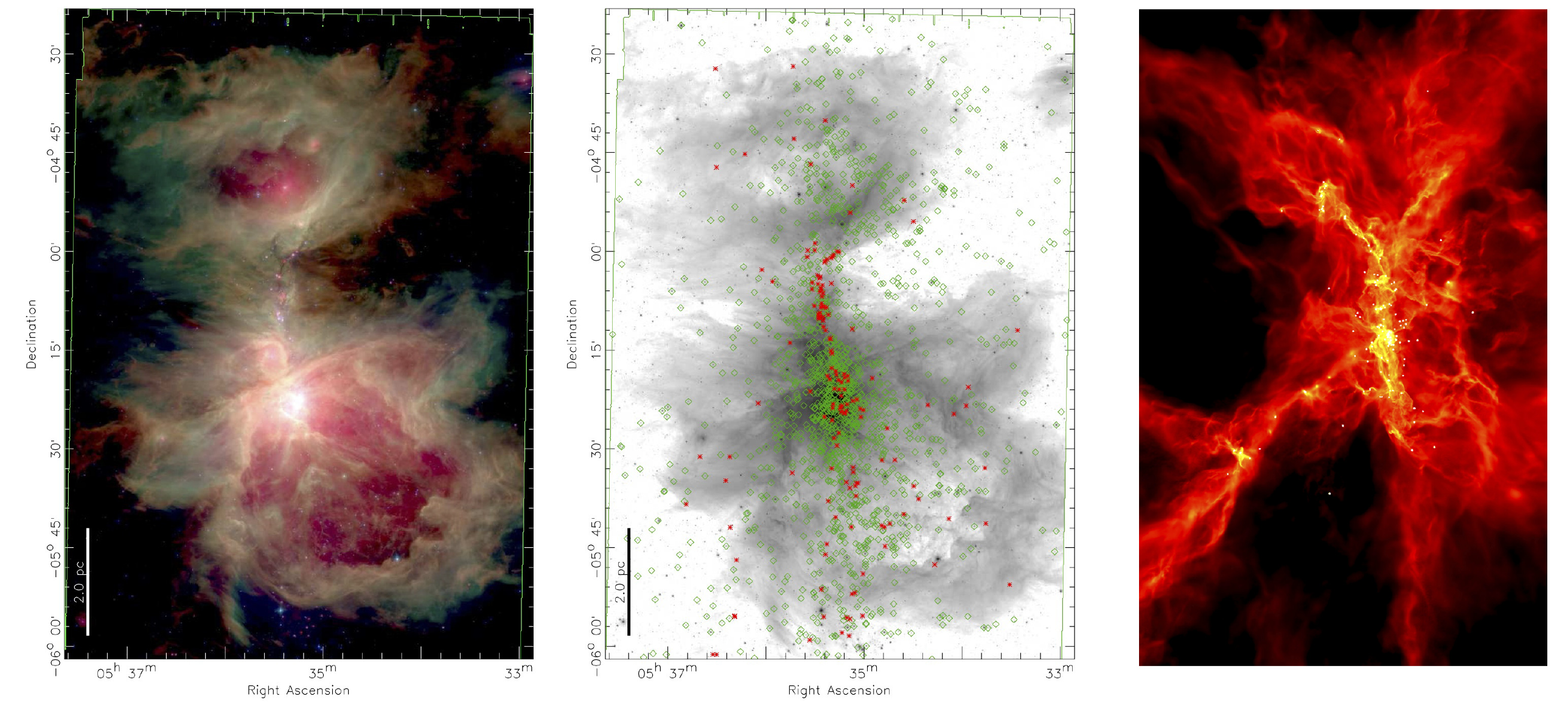}
\label{orion-bate}
\caption{\it \small Comparison between the results of an SPH simulation of formation of a stellar cluster \citep[right,][]{bate09} and mid-infrared observations of the Orion Nebula Cluster \citep[left, R/G/B = 24~$\mu$m / 5.8~$\mu$m / 4.5~$\mu$m,][]{mege12} and the distribution of stars formed within the region \citep[centre, Class~II young stellar objects shown in green, Class~I protostars shown in red,][]{mege12}. Image from \citet[][left and centre panels]{mege12}, reproduced by permission of the AAS.}
\end{center}
\end{figure}

The first contributed talk of this session was given by Matthew Bate (University of Exeter) who presented the kinematic properties of protostellar groups and clusters produced from numerical simulations of star formation. Such simulations are very dynamic and chaotic, a picture which is not often welcomed by observational astronomers because of the difficulty relating the static observational picture of star formation with the underlying highly dynamical physical processes. For example in a typical young star cluster that is a few Myrs old and several parsecs across a star moving at a velocity of 1~km/s ($\simeq$ 1~pc/Myr) can have moved a significant distance in its short lifetime relative to the size of the region. Analysing such regions as if they are completely static and assigning labels to stars such as ``clustered'' and ``isolated'' is therefore risky. Despite this these simulations have been able to reproduce many of the observational properties of young stars and clusters, from the mass function of stars and brown dwarfs to the frequency and properties of binary stars \citep[see Figure~5 and][]{bate12}.

Matthew also presented results obtained by coupling hydrodynamical simulations of star formation with N-body simulations of the cluster dynamics that follow once all the stars have formed \citep{moec10}. Their results show that clusters can form very dense, with half-mass radii of 0.05~pc, but they can then expand to $\sim$1~pc, similar to the size of the Orion Nebula Cluster, and containing 30-40\% of the stellar mass, with the remaining stars occupying a halo that freely expands to $>$100~pc in $<$10~Myr. The resulting stellar velocity dispersion, driven by dynamical interactions, is often greater than that of the dense cores from which the stars formed \citep{bate03} and increases in the halos of star clusters, though there is little dependence of the velocity dispersion on stellar mass. Numerical simulations such as these make many predictions that can be tested using kinematics from Gaia, including how the velocity dispersions of stars varies as a function of stellar mass and environment, and the expansion and evolution of star clusters that can be studied using proper motions.

The final contributed talk of the day was given by Emilio Alfaro (Instituto de Astrofisica de Andalucia). Emilio reminded us that star formation is one of the most challenging problems in modern astronomy because of the wide range of spatial scales over which multiple different physical processes take place. This is further complicated by our inability to observe the full 6-dimensional phase space structure of the evolving stellar systems we wish to describe, with most observational studies limited to only 3-dimensional information (2 positional dimensions and radial velocities).

Information in the two positional dimensions has been used by many people to quantify the structure of star forming regions and star clusters, and to understand how this structure evolves over time \citep[e.g.,][]{cart04,sanc09}. To better understand this evolution we also need to study what is happening in the kinematic subspace during this evolution. There have already been a number of detections of kinematic substructure \citep[e.g.,][see Figure~7]{fure06,jeff14}, suggesting it might be a very common phenomenon in young star forming structures. Emilio explained how this kinematic substructure can be detected and measured using the minimum spanning tree, an algorithm that has already been effective in diagnosing spatial substructure \citep{cart04}. Emilio then showed how this method could be applied to existing radial velocities for the star forming region NGC~2264, revealing the presence of three phase-space groupings in the region.

There is currently a serious lack of statistical tools for analysing the kinematics of star forming regions and star clusters and performing pattern analysis in velocity subspace. Work such as this offers real potential for overcoming this deficit by providing new tools for comparing observations and the predictions of simulations. One particular strength of this method is that is can be easily implemented and leads to a quantitative description of the kinematic structure that will allow effective comparisons between different clusters, environments and datasets in a homogeneous way.

\begin{figure}[t]
\begin{center}
\includegraphics[width=17cm]{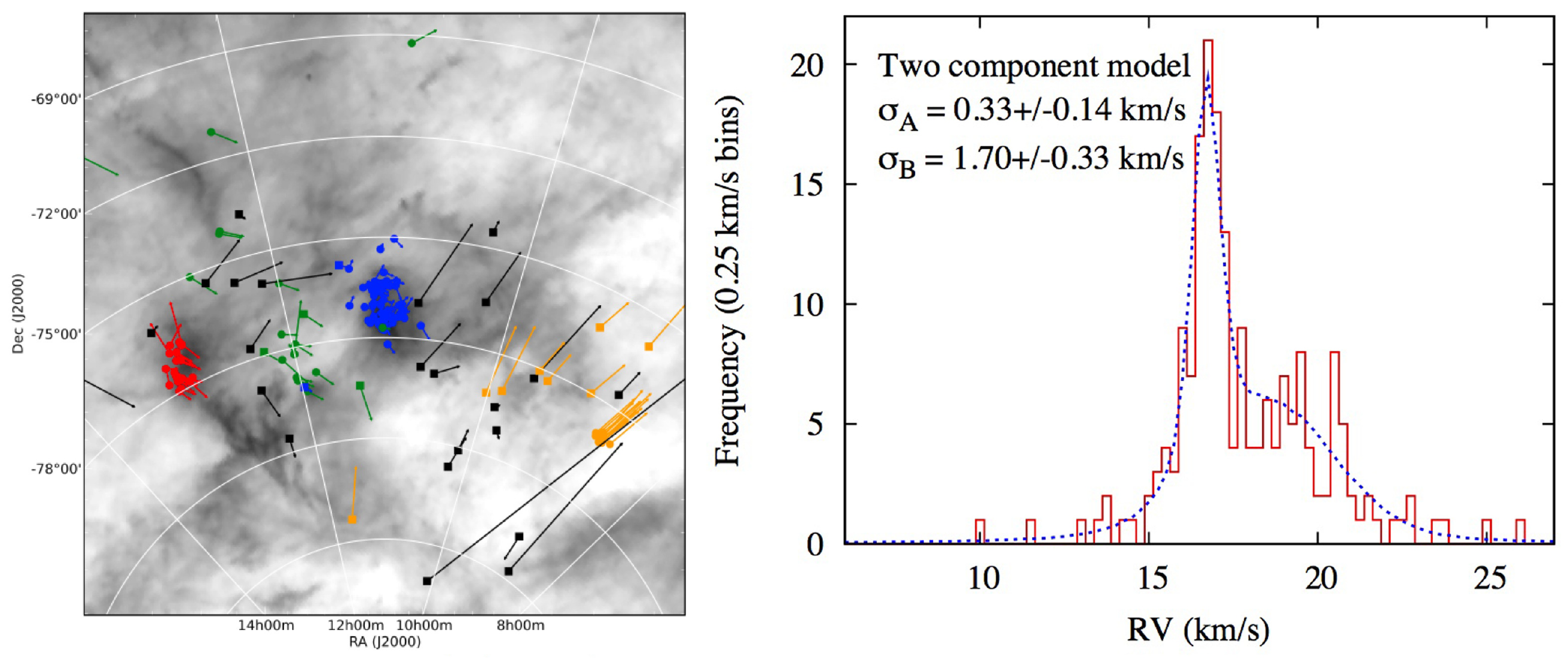}
\label{kinematic-substructure}
\caption{\it \small Left: The kinematics of the Chamaeleon star forming region derived from proper motions (motions for $10^5$~yrs shown) projected against a dust map \citep{lope13}. Shown are members of the Cha~I (blue), Cha~II (red), $\eta$~Cha (yellow), and $\epsilon$~Cha (green) groups, as well as background sources (black). Right: Radial velocity distribution for members of the $\gamma^2$~Vel association \citep{jeff14} showing the best fit model using two Gaussian components representing a gravitationally bound cluster (component A) and a slightly younger, gravitationally unbound OB association (component B).}
\end{center}
\end{figure}

The final invited talk of the day was given by Estelle Moraux (University of Grenoble) who reviewed the current state of kinematic observations of young stars and clusters. She reminded us that kinematics provide one of the key observations to understand star formation in clusters and the evolution of this clustering. There are many ways that kinematic observations facilitate this, including improving the membership of clusters, tracing back the motions of stars, understanding the dynamical state of clusters, and identifying kinematic substructures. This is the key to understanding not only the star formation process but also the formation and evolution of star clusters.

Kinematic measurements are an important discriminant for assessing the membership of star clusters, OB associations, and young moving groups and to understand their origins. For example, \citet{lope13} used proper motions to show the Cha~I and Cha~II moving groups are connected to each other, but that both appear to be unrelated to the $\epsilon$ and $\eta$~Cha groups (see Figure~7). Radial velocities are also useful for identifying cluster members \citep{bric07a} and there are many methods that can be used to exploit these measurements to resolve cluster membership (e.g., the minimization of kinematic distances or the convergent point method). For older moving groups this method can become difficult because the resonant trap can significantly increase the number of contaminants \citep{fama08}, so additional criteria such as chemical composition are needed.

Kinematic observations are fundamental for providing a measure of the dynamical state of a cluster or OB association, either using proper motions \citep[e.g.,][]{pang13} or radial velocities \citep{cott12}, though the latter need to be corrected for the broadening of the velocity dispersion due to binaries. Such measurements can provide an indication of whether the regions are in virial equilibrium or in the process of collapsing or expanding, and can also reveal the presence of cluster rotation \citep{hena12} and kinematic substructure \citep[][see Figure~7]{fure06,jeff14}. Proper motions and radial velocities can also be used to measure kinematic parallaxes, which \citet{gall13} used to reveal significant depth effects between the different subgroups of the Lupus associations, showing that the (likely older) weak-lined T-Tauri stars are more dispersed in both angular extent and depth compared to the (younger) Classical T-Tauri stars.

While kinematics has already taught us a considerable amount about cluster membership, their global dynamical states, as well as their internal kinematics, very few regions have been studied in enough detail to build up a coherent picture of the dynamical evolution of young stars. We need to study both more clusters and associations, but also obtain higher precision kinematic measurements for stars in these regions, ideally better than 1~km/s. Gaia will provide high-precision proper motions down to $G = 20$~mag, but its radial velocities will not do better than 1~km/s, even for the brightest stars, and won't provide any accuracy beyond $G = 12$~mag. To maximise the potential of Gaia complementary data are needed, including high-prevision radial velocities (e.g.,  from the Gaia-ESO Survey or future facilities such as WHT/WEAVE or VISTA/4MOST) and complementary, deeper proper motions (e.g., from Pann-STARRS and LSST).

To deal with this wealth of data from both Gaia and complementary programmes we need to be prepared. Hydrodynamic and N-body simulations of star forming regions and star clusters are needed to compare to observations and to link the collapse of molecular clouds with the dynamical evolution of young star clusters. Analysing this data will also provide new challenges, requiring observers to study young stars and star clusters in 6-dimensional phase space (three position and three velocity dimensions), as well as folding in a wealth of multi-dimensional data. This will require new statistical tools to efficiently analyse the observations and perform quantitative comparisons with the results of simulations.

\section*{How can we best prepare for Gaia data?}

The meeting ended with a look ahead to when the Gaia data will become available, and in what form it will take. The first data release, planned for the Summer of 2016, will include early photometry and proper motions for stars detected by both Hipparcos and Gaia, from which the $\sim$25~year baseline will allow proper motions with precisions up to $\sim$50~$\mu$as to be calculated. Following that it is envisioned that yearly data releases will be made from 2017 onwards, potentially including photometry, astrometric solutions, radial velocities, and orbital solutions for binary systems. This will culminate in a full and final data release scheduled for 2022.

In the discussion session that followed it was noted that while some major developments will be possible using data from the first release (2016), many will require the improved proper motions and parallaxes that come with later data releases (2017--2022). For example, it should be possible to resolve the origins of OB associations by tracing back the motions of their members to see if they represent an expanded star cluster. For simple systems this may be possible with early Gaia data, though more complex arrangements of multiple expanding star clusters will require parallaxes and more precise proper motions from later data releases.

One issue that was agreed upon by all attendees was that we will need many N-body simulations that cover a wide variety of initial and final conditions so that we can clearly compare observations with simulations. We will also need to develop a number of new diagnostics that will allow us to quantitatively compare observations and simulations. These diagnostics will need to be extensively tested against N-body simulations to make sure that they reliably trace the evolution of the cluster and can be used to resolve the initial conditions. If they are to be effectively used to constrain the initial conditions of star clusters then these diagnostics also need to be resistant to the motions of the stars as the region evolves. To fully exploit the immense amount of data that Gaia will provide we need to develop and test these tools as soon as possible.

\section*{Final words}

The study of the star formation process has made considerable progress in the last half century, thanks to a wide variety of observatories, instruments and techniques. Over the last decade, observational studies of star forming regions have mainly focussed on identifying objects and building up an understanding of the evolutionary sequences followed by forming stars. We are now entering the era of kinematics, thanks both to Gaia and many ground-based radial velocity surveys. These should allow great progress to be made in understanding the dynamical evolution of star-forming structures and will allow us to link our current `observational snapshots' to the evolution of these regions predicted by numerical simulations.

The take-home message from this meeting was that Gaia will play a vital role in improving our understanding of both the star formation process and the formation and evolution of star clusters, but it won't tell us everything. Complementary spectroscopic information, including spectral types and radial velocities, will maximise the impact of the data, while N-body simulations and diagnostic tools will allow us to efficiently compare observations and simulations. With these tools the full potential offered by Gaia can be realised, providing a revolution in our understanding of the star and cluster formation processes.

\vspace{1cm}

\noindent
{\it \small The RAS Specialist Discussion Meeting on 9 January 2015 was organised by Nicholas Wright, Ernest Rutherford Fellow at Keele University, Richard Parker, RAS research fellow at Liverpool John Moores University, and Rowan Smith, Norman Lockyer Fellow at the University of Manchester. The author would like to thank the Royal Astronomical Society for helping organise the meeting, as well as all those who contributed to and helped organise the meeting, both for making their presentation materials available and for giving comments on this article.}

\bibliographystyle{mn2e}
\bibliography{/Users/nwright/Documents/Work/tex_papers/bibliography.bib}

\end{document}